\documentclass[]{IEEEtran}
\IEEEoverridecommandlockouts
\newcommand{\ignore}[1]{}
\usepackage{subcaption}%
\usepackage{graphicx}
\usepackage{lipsum}
\usepackage{textcomp}
\usepackage{xcolor}
\usepackage{float}
\usepackage{multirow}
\usepackage{adjustbox}
\usepackage{xspace}
\usepackage{pifont}
\usepackage{tikz,hyperref}
\usepackage{amsmath}

\usepackage{tikz,xcolor,hyperref}

\definecolor{lime}{HTML}{A6CE39}
\DeclareRobustCommand{\orcidicon}{%
	\begin{tikzpicture}
	\draw[lime, fill=lime] (0,0) 
	circle [radius=0.16] 
	node[white] {{\fontfamily{qag}\selectfont \tiny ID}};
	\draw[white, fill=white] (-0.0625,0.095) 
	circle [radius=0.007];
	\end{tikzpicture}
	\hspace{-2mm}
}

\foreach \x in {A, ..., Z}{%
	\expandafter\xdef\csname orcid\x\endcsname{\noexpand\href{https://orcid.org/\csname orcidauthor\x\endcsname}{\noexpand\orcidicon}}
}

\usepackage{setspace}

\hypersetup{
  colorlinks,
  allcolors=blue
  citebordercolor=blue,
  citecolor=blue,
  linkcolor=blue,
  urlcolor=blue}

\newcommand{\Design}{CFUNG\xspace}
\usepackage{todonotes}

\begin{document}
\title{Power- and Area-Efficient Unary Sorting Architecture Using FSM-Based Unary Number Generator}

\author{Amir Hossein Jalilvand$^*$ \orcidA and M. Hassan Najafi$^+$ \orcidC
\\
$^*$Electrical and Computer Engineering, University of Louisiana at Lafayette, LA, USA\\$^+$Electrical, Computer, and Systems Engineering Department, Case Western Reserve University, OH, USA
\vspace{-1em}
}

\maketitle
\begin{abstract}
Sorting is a fundamental operation in computer systems and is widely used in applications such as databases, data analytics, and hardware accelerators. Unary computing has recently emerged as a low-cost and power-efficient paradigm for implementing hardware sorters by eliminating the need for complex arithmetic operations. However, existing comparison-free unary computing-based designs suffer from significant area and power overhead due to costly unary number generators.

In this paper, we present a novel ascending-order unary sorting module featuring a finite-state-machine-based unary number generator that significantly reduces implementation costs. By generating right-aligned unary streams using a two-state finite-state machine, our architecture iteratively identifies the minimum input value in each cycle without conventional comparators. Synthesis results in a 45nm technology node demonstrate up to \textbf{82\% reduction in area} and \textbf{70\% reduction in power consumption} compared to state-of-the-art unary designs. The proposed sorter offers a promising solution for energy-constrained and resource-limited hardware systems.
\end{abstract}


\section{Introduction}
Sorting is a fundamental operation in many domains, including databases, data analytics, and computer algorithms. Numerous software- and hardware-based sorting techniques have been developed to improve sorting speed and efficiency~\cite{jalilvand2025sorting}. While both approaches aim to organize data effectively, they differ significantly in methodology~\cite{10472626}.
In software-based sorting, the sequence of comparisons and data exchanges is determined by the specific data values. In contrast, hardware-based sorting is typically designed to operate independently of input data values. As a result, hardware implementations primarily focus on minimizing area overhead and power consumption, which are critical metrics in energy-constrained systems.
This concern becomes even more pronounced in embedded systems, where chip area and power availability are severely limited. Furthermore, as fabrication technologies continue to scale, managing thermal dissipation becomes increasingly important to counteract the exponential rise in leakage current with temperature. Consequently, designing sorting architectures that are both area- and power-efficient is essential for practical deployment.
Hardware-based sorting solutions are generally classified into two categories ~\cite{jalilvand2025sorting}:
\begin{enumerate}
    \item \textbf{Comparison-based sorting}, where elements are explicitly compared and rearranged, such as in Batcher networks;
    \item \textbf{Comparison-free sorting}, which avoids explicit data comparisons and instead relies on structural or algorithmic techniques to achieve ordered output.
\end{enumerate}

\begin{figure}[t!] 
	\centering
	\includegraphics[width=2.7in]{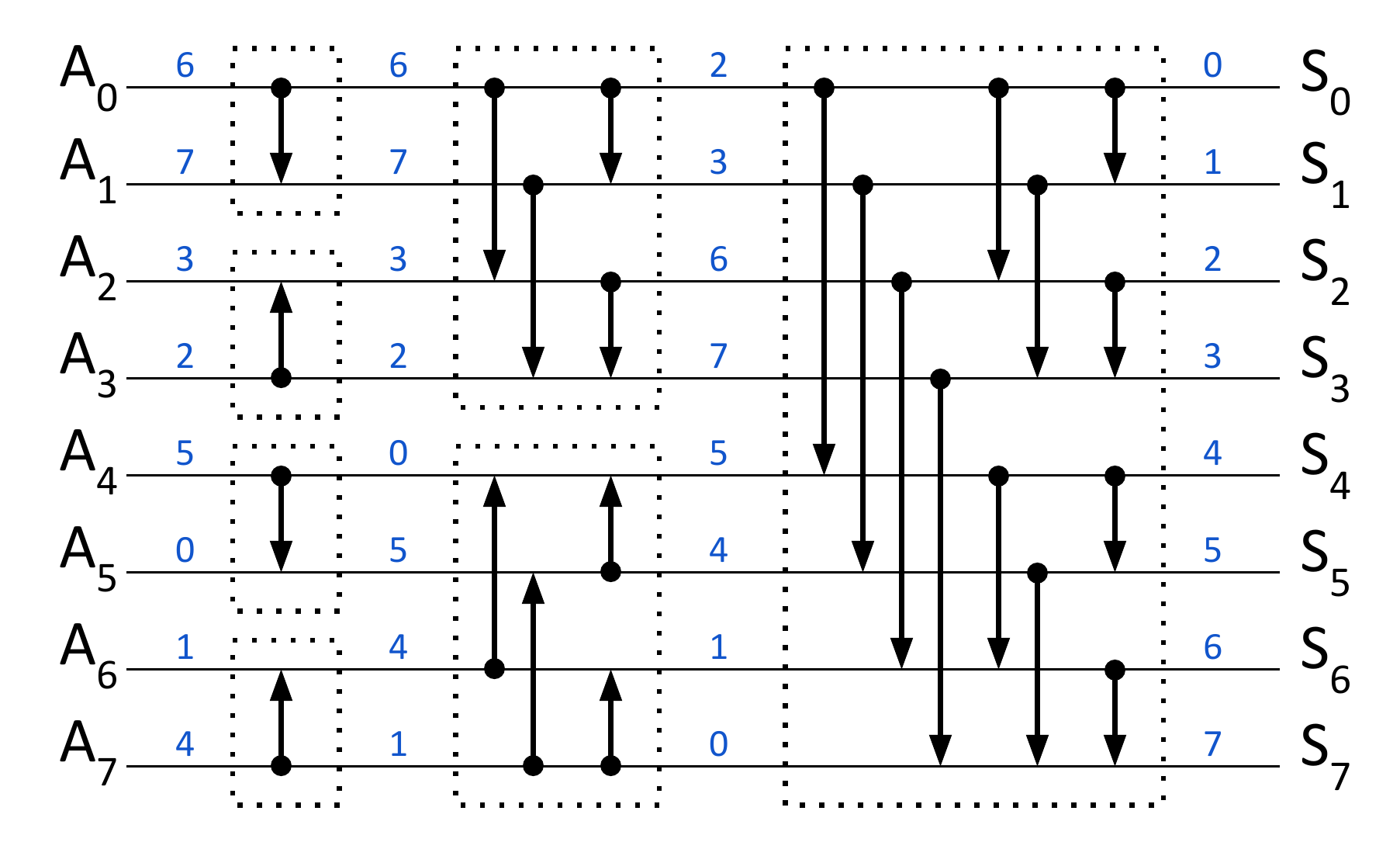}
	
	\caption{The CAS network for sorting 8 input data
	~\cite{Najafi2018Sorting}.}
	\label{Bitonic8}
	\vspace{-1.em}
\end{figure}

\begin{figure}[t!]
	\centering
	\includegraphics[trim={0.5cm 0.5cm 0.5cm 0.5cm },clip,width=0.85\linewidth]{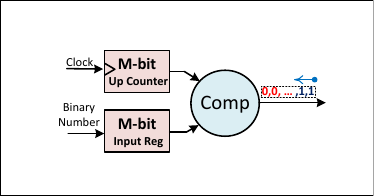}
	\caption{Conventional Unary Bit-stream Generator. The circuit generates a right/left aligned bit-stream with an up/down counter.}
	\vspace{-1em}
	\label{FigCovUng}
\vspace{-.95em}	
\end{figure}

Batcher's sorting is a notable example of comparison-based sorting~\cite{Batcher1968}. The sorting is performed 
 by wiring a network of compare-and-swap (CAS) units.
 Each CAS block compares two input values and swaps the values at the output, if required. There are two variants of CAS blocks: an "ascending ($\uparrow$)" type and a "descending ($\downarrow$)" type. Fig~\ref{Bitonic8} shows the CAS network for sorting eight input data. 
 The hardware cost and the power consumption of Batcher's network depend on the number of CAS blocks and the cost of each CAS block. 
 The total number of CAS blocks in an $N$-input Batcher's sorting is $N \times log_2(N) \times (log_2(N)+1)/4$. Thus, 8-, 16-, 32-, and 256-input Batcher networks require 24, 80, 240, and 4,608 CAS blocks, respectively \cite{farmahini2012modular}. 
Batcher's sorting is conventionally implemented based on the weighted binary representation. Binary representation is compact; however, computation on this representation is relatively complex. The complexity 
increases by increasing the data-width. Increasing the complexity 
affects the cost of hardware implementation, latency, power, and hence, 
energy consumption. 
Najafi \textit{et al.} proposed an alternative low-cost hardware design for Batcher's networks using \textit{unary computing} (UC)~\cite{Najafi2018Sorting}. 

The minimum and the maximum value functions, the essential functions in building Batcher sorting networks, can be realized efficiently in the unary domain using simple bit-wise \texttt{AND} and \texttt{OR} operations. An area and power saving of up to 92\%
is reported in~\cite{Najafi2018Sorting} for the unary Batcher sorting design compared to the conventional binary counterpart.

\begin{figure*}[t]
     \centering
    
     \begin{subfigure}[]{0.49\textwidth}
         \centering
         \vspace{0.75em}
         \includegraphics[trim={0.5cm 0.8cm 0.5cm 0.5cm },clip, height=0.21\textheight]{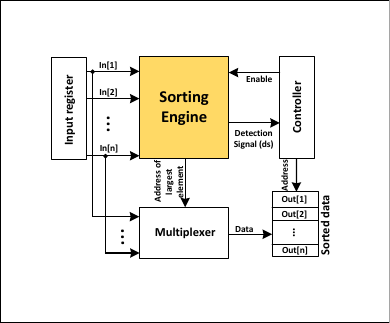}
	\vspace{1.2em}
 	\caption{}

	\label{ToplevelDAC2022}
     \end{subfigure}
     \begin{subfigure}[]{ 0.49\textwidth}
         \includegraphics[trim={0.5cm 0.8cm 0.8cm 0.5cm},clip, width=0.95\textwidth]{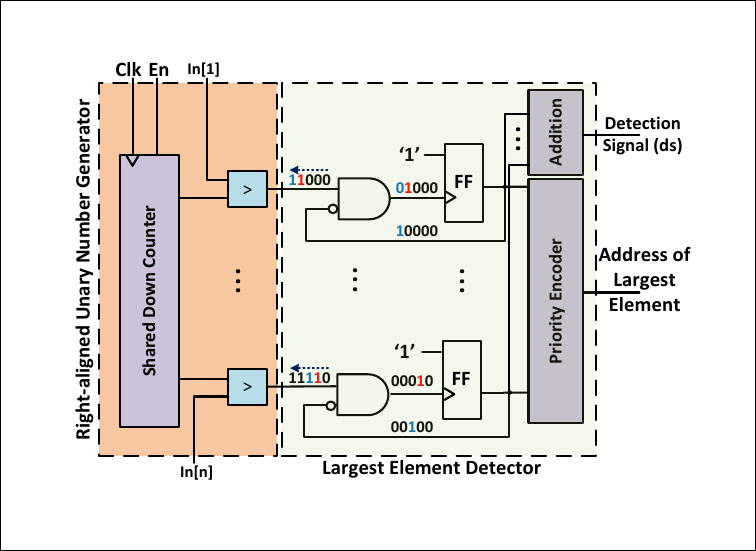}
	\vspace{-0.7em}
	\caption{}
	\label{SEDAC2022}
     \end{subfigure}
\vspace{-0.7em}
\caption{Architecture of  previous comparison-free unary sorter in ~\cite{AmirDAC2022}. \ref{ToplevelDAC2022}: general architecture,  \ref{SEDAC2022}: sorting engine. } 

        \label{fig2:PreviousWork}
\vspace{-1.5em}        
\end{figure*}

The hardware design of a comparison-free sorting module is proposed in~\cite{Ghosh2019}. Their design sorts $N$ data elements in nearly $N$ clock cycles while recognizing 
the maximum number in the 1st clock cycle. Their sorting module is constructed by employing $N$ symmetric cascaded blocks, and sorting operations are performed in a pipelined fashion.
A comparison-free sorting algorithm is also introduced in~\cite{Abdel-Hafeez2017An}. This design can be applied to any data distribution with no significant adjustment. The number of required cycles falls in the range of $2N$ to $2N + 2K - 1$, where $K$ is the bit-width of data 
and $N$ is the number of input data.
Recently,  in~\cite{AmirDAC2022} we
proposed a descending-order comparison-free sorting architecture based on UC.
Motivated by the idea of finding the 
1st 
one in the left-aligned unary bit-stream, which addresses the index of the maximum value in the unary processing domain~\cite{Jalilvand2020Fuzzy-Logic},  our design 
iteratively finds the index of the maximum value by converting data to left-aligned unary bit-streams and finding the first ``1'' in the generated bit-streams. 
The proposed sorter sorts many inputs in fewer clock cycles compared to the unary design of~\cite{Najafi2018Sorting}. 
Synthesis results reveal up to 60\% savings in hardware area footprint compared to the state-of-the-art designs.
The major 
drawback of this 
design is the cost of unary bit-stream generation. 
Converting input data 
from weighed binary radix to unary bit-streams 
with current comparator-based unary bit-stream generators (Fig.~\ref{FigCovUng}) is costly in terms of area and power. This issue further aggravates when the number of inputs and the data precision increase. In this work, we improve our previous 
work and 
propose a low-cost UC-based 
sorting module with modified unary number generator (UNG), which unlike our work in \cite{AmirDAC2022}, 
iteratively finds the minimum values and sorts the input values in ascending order. To reduce the cost of bit-stream generation, the 
new design utilizes a finite-state-machine (FSM)-based 
UNG called \Design (Comparison Free UNG) ~\cite{banitaba2025cfung}, which can generate 
right-aligned unary bit-streams. 
Synthesis results show up to 82\% and 70\% savings in the area and power of the 
proposed design compared to the our state-of-the-art work in ~\cite{AmirDAC2022}.

\ignore{
\begin{figure}[]
	\centering
	\includegraphics[trim={0.5cm 0.5cm 0.5cm 0.5cm },clip,width=0.85\linewidth]{Fig1.pdf}
		\vspace{-1.25em}
	\caption{Conventional Unary Bit-stream Generator. The circuit generate a right/left aligned bit-stream with an up/down counter.}
	\vspace{-1em}
	\label{FigCovUng}
\end{figure}
}
The rest of the paper is organized as follows. 
In section ~\ref{Previous}, we briefly review the idea of our previous work in \cite{AmirDAC2022}. In section ~\ref{Proposed}, we propose our 
ascending-order sorting unit based on \Design generator. 
Section~\ref{Design_Eval} presents design evaluation results. 
Finally, Section~\ref {Conclusion} concludes the paper. 

\section{Comparison-free unary sorter with traditional UNG}

\label{Previous}

\begin{figure}[]
	\centering
\vspace{-0.5em}
	\includegraphics[trim={1cm 1.5cm 1.5cm 1.3cm },clip,width=0.5\linewidth]{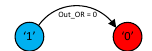}
	\vspace{-0.5em}
	\caption{FSM representation of the \Design module used for right-aligned unary number generation. The FSM operates with two states and produces a unary bitstream where the number of logic-‘1’ bits corresponds to the input binary value.}
	\label{fugenFSM}
\end{figure}

In this section, we briefly introduce our previous idea of a comparison-free unary sorter and discuss the significant area-power bottleneck of this design. Fig.~\ref{fig2:PreviousWork} shows the 
architecture of our previous comparison-free unary sorter in ~\cite{AmirDAC2022}. It consists of two main parts: \ref{ToplevelDAC2022}: general architecture,  \ref{SEDAC2022}: sorting engine.
As shown in Fig.~\ref{ToplevelDAC2022},  the architecture consists of three main components: a sorting module, a controller, and a multiplexer. The design begins by reading unsorted data from input registers. The sorting module then performs the sorting process by identifying the address of the maximum number at each step. The controller manages the overall operation of the sorting process, coordinating the actions of the sorting module. Finally, the multiplexer handles the selection and routing of data within the architecture. Together, these components enable the sorting of the input data by iteratively identifying the maximum number and organizing the data accordingly.
The sorting module in this architecture (Fig.~\ref{SEDAC2022}) converts the data into right-aligned unary bit-streams. It then determines the index of the bit-stream that corresponds to the maximum value. This is achieved by identifying the bit-stream that produces the first occurrence of a 1, indicating the maximum value in the data set. By performing this operation, the sorting module can efficiently identify and track the maximum value throughout the sorting process. 

Consider each input is converted into a corresponding right-aligned unary bit-stream. The sorting process begins by counting down in cycles. In each cycle, the system generates specific bits for the inputs based on their unary representations.
During the process, the flip-flops corresponding to the inputs with the maximum values are activated. The detection signal (\texttt{ds}), which is obtained by adding the activated bits, indicates the number of inputs with the maximum value. If ds is greater than one, it means there are multiple inputs with the same maximum value.
To determine the memory address of one of the maximum values, a priority encoder is employed. This encoder selects one of the inputs with the maximum value based on a predefined priority scheme. The value of ds is then passed to the controller, which controls the further operations of the sorting process. 
When the detection signal (\texttt{ds}) equals 2, indicating that there are two numbers with the maximum value, the state of the controller's state machine changes from "Find the index" to "Put the results."

\begin{figure*}[t]
     \centering
    
     \begin{subfigure}[]{0.45\textwidth}
         \centering
         \vspace{2.5em}
        \includegraphics[trim={0.75cm 0.75cm 0.75cm 0.75cm },clip,width=0.9\textwidth]{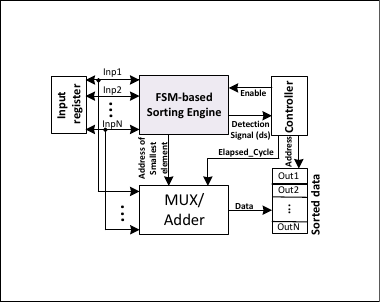}
	\vspace{2.2em}
 \caption{}
	\label{ToplevelNEW}
     \end{subfigure}
     \begin{subfigure}[]{ 0.54\textwidth}
     \centering
         \includegraphics[trim={0.75cm 0.75cm 0.75cm 0.5cm },clip,width=\textwidth]{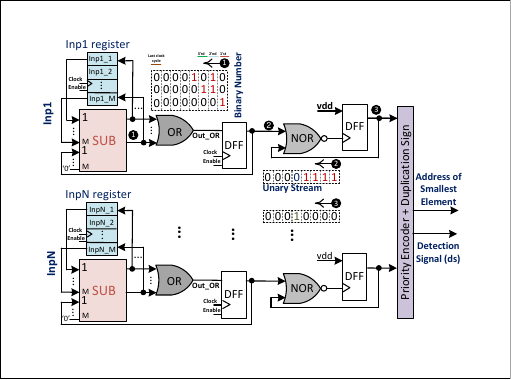}
 \vspace{-1.5em}
 \caption{}
	\label{SEThiswork}
     \end{subfigure}
\vspace{-0.7em}
\caption{Architecture of our proposed comparison-free unary sorter with modified UNG. \ref{ToplevelNEW}: general architecture,  \ref{SEThiswork}: sorting engine. } 

        \label{Fig3:NewWork}
\end{figure*}

\begin{table}[]
\caption{\small Area and power cost of converting eight input data with data precision $M$ to corresponding unary bit-streams. 
}
\vspace{-0.25em}
\centering
\label{Tbl:eightinput}
\renewcommand{\arraystretch}{.9}
\resizebox{3.2in}{!}{
\begin{tabular}{|c|cc|cc|}
\hline
\multirow{2}{*}{M} & \multicolumn{2}{c|}{\begin{tabular}[c]{@{}c@{}}Area ($\mu m^2$)\end{tabular}} & \multicolumn{2}{c|}{\begin{tabular}[c]{@{}c@{}}Power @$100$MHZ ($\mu W$)\end{tabular}} \\ \cline{2-5} 
                   & \multicolumn{1}{c|}{Comp. Based}                    & \Design                   & \multicolumn{1}{c|}{Comp. Based}                         & \Design                          \\ \hline
8                  & \multicolumn{1}{c|}{709}                                 & 210                     & \multicolumn{1}{c|}{42.6}                                   & 17.7                         \\ \hline
16                 & \multicolumn{1}{c|}{1,564}                               & 398                     & \multicolumn{1}{c|}{71.0}                                   & 18.0                         \\ \hline
32                 & \multicolumn{1}{c|}{3,560}                               & 680                     & \multicolumn{1}{c|}{130.3}                                  & 22.1                         \\ \hline
\end{tabular}
}
\vspace{-1em}
\end{table}

\section{Comparison-free unary sorter with modified UNG}
\label{Proposed}

UC enables highly simplified and low-power hardware implementations for fundamental operations such as sorting. However, a major bottleneck in existing comparison-free unary sorting architectures is the overhead associated with unary number generation. Traditional UNGs, typically implemented using counters and comparators, introduce significant area and power costs, particularly as input size and data precision increase.

To address this limitation, we propose an enhanced comparison-free unary sorting architecture that incorporates a finite-state-machine-based unary number generator (FSM-based UNG), referred to as CFUNG. The CFUNG unit replaces the conventional comparator-based circuitry with a compact two-state FSM capable of generating right-aligned unary bitstreams. This improvement not only reduces the complexity of unary stream generation but also significantly lowers the overall hardware cost.

The modified sorting architecture retains the core principles of unary processing while introducing several key enhancements:
\begin{itemize}
    \item A restructured sorting engine capable of identifying the \textit{minimum} values (instead of maximum as in previous designs), enabling ascending-order sorting;
    \item Integration of a low-cost CFUNG unit for efficient bitstream generation;
    \item A lightweight detection and control mechanism that tracks the position of the current minimum value using a priority encoder and a finite-state controller.
\end{itemize}

\subsection{Novel UNG UNIT}

Asadi \textit{et al.} ~\cite{sinaDate2021} introduced a low-cost LD bit-stream generator based on FSM, eliminating the requirement for a comparator unit. This innovative generator achieved a significant reduction of up to 80\% in hardware area and area-delay product compared to the leading comparator-based LD bit-stream generator, all while ensuring the precision of the generated bit-streams. It's worth noting, though, that their design does not support the creation of a unary stream, which is a sequence of consecutive '1's followed by a sequence of consecutive '0's and vice versa.
Compared to the conventional approach of generating unary bit-streams using a traditional right-aligned unary bit-stream generator which is based on  
a pair of counter and comparator, in this work, we utilize \Design that reduces the area and power consumption significantly. It generates right-aligned bit-streams 
and is based on a two-state FSM, as 
illustrated in Fig.~\ref{fugenFSM}.
Table~\ref{Tbl:eightinput}
compares the area and power cost of converting eight input data to their corresponding right-aligned bit-streams. 
As 
seen, \Design  reduces the power and area consumption up to $83\%$ and $81\%$, respectively.

\ignore{
(Fig \ref{Controller_DAC2022})  In this state, the controller waits until both numbers are present in the output. It does not change the state until this condition is met.
Once both numbers are available in the output, the controller activates a down counter. This down counter provides the memory address of the numbers found during the "Put the results" state. Additionally, a multiplexer is utilized to obtain the address of the maximum values from the sorting module. The multiplexer also retrieves the weighted-binary representation of the maximum value, which is then stored in the output registers. 
}

\ignore{
\begin{figure}[]
	\centering
	\includegraphics[trim={0.85cm 1cm 0.8cm 0.9cm },clip,width=0.60\linewidth, 
	]{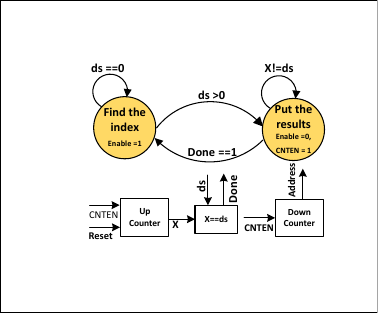}
	\vspace{-1em}
	\caption{Controller of the first design.}
	\label{Controller_DAC2022}
\vspace{-1em}
\end{figure}
}

\begin{figure}[]
\vspace{-0.25em}
 \centering
	\includegraphics[trim={1cm 0.8cm 3cm 1cm },clip,width=0.7\linewidth]{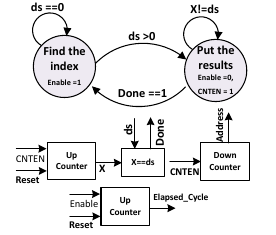}
	\vspace{-0.5em}
	\caption{Controller of the second design.}
	\label{Controller}
\vspace{-2em}
\end{figure}


\subsection{Proposed Sorting Unit}

Fig.~\ref{ToplevelNEW} shows the high-level architecture  
includes an FSM-based sorting module, a controller, a multiplexer (MUX), and an adder. The design reads unsorted data from the input registers and performs sorting 
by finding the address of the minimum number at each step. Fig.~\ref{SEThiswork} shows the proposed  FSM-based sorting module, consisting of two main components:  a right-aligned unary bit-stream generator and the smallest element detector (SED).

 As can be seen, the FSM has only one transition from a ``blue" state to a ``red" state.  When the FSM is in the ``blue" state,  the output produces 
'$1$'. 
The output then becomes
 '$0$' when the state 
 changes to ``red".  
The FSM goes from the ``blue" state to the ``red" state when the \texttt{Out\_OR} signal becomes '$0$'.
For example, consider converting 
$p_1 = 0.5$ with \Design to a right-aligned unary bit-stream 
with a bit-width ($M$) of $3$ (bit-stream length of 8). 
In the first cycle, the binary value of $p_1$ 
($0.100$) is subtracted from $0.000$, and the result becomes $0.100$, as shown in Fig~\ref{SEThiswork}-\ding{202}. The new value 
of \texttt{Out\_OR} signal ('1') is sent 
out when the clock rises. The generated 
bit-stream is shown 
in \ding{203}.
In the next cycle, 
$0.100$ is subtracted from the previously generated bit (here, '1'). 
Therefore, $0.100$ is subtracted from $0.001$ and the output of subtraction becomes $0.011$. Again, the \texttt{Out\_OR} signal is '1'. This process finally produces 
the right-aligned bit-stream $00001111$, as 
shown in \ding{203}. 

 \ignore{
    \begin{figure}[t]
	\centering
	\includegraphics[trim={0.5cm 0.6cm 0.5cm 1.1cm },clip,width=0.80\linewidth, 
	]{FigS2}
	\vspace{-2em}
	\caption{High-level architecture of the second design.}
	\label{GeneralArchitecture}
\end{figure}
}

\begin{table*}[]
\centering
\caption{\small Hardware Cost Comparison of 
Sorting $N$ inputs with 
Different Data Bit-widths $M$.}
\label{tbl:2}
\begin{tabular}{|c|c|lccc|cccc|cccc|}
\hline
\multirow{2}{*}{N}   & \multirow{2}{*}{M} & \multicolumn{4}{c|}{Area ($\mu m^2$)}                                                                                       & \multicolumn{4}{c|}{Power ($mW$) @ max freq}                                                                    & \multicolumn{4}{c|}{Critical Path  ($ns$)}                                                                                     \\ \cline{3-14} 
                     &                    & \multicolumn{1}{c|}{\cite{Ghosh2019}}    & \multicolumn{1}{c|}{\cite{Najafi2018Sorting}}   & \multicolumn{1}{c|}{~\cite{AmirDAC2022}} & Prop. & \multicolumn{1}{c|}{\cite{Ghosh2019}} & \multicolumn{1}{c|}{\cite{Najafi2018Sorting}} & \multicolumn{1}{c|}{~\cite{AmirDAC2022}} & Prop. & \multicolumn{1}{c|}{\cite{Ghosh2019}}  & \multicolumn{1}{c|}{\cite{Najafi2018Sorting}} & \multicolumn{1}{c|}{~\cite{AmirDAC2022}} & Prop. \\ \hline
                  & 8                   & \multicolumn{1}{l|}{1,752}   & \multicolumn{1}{c|}{2,194}   & \multicolumn{1}{c|}{1,308}        & 918           & \multicolumn{1}{c|}{ 3.23} & \multicolumn{1}{c|}{3.3}   & \multicolumn{1}{c|}{5.2}          & 3.48          & \multicolumn{1}{c|}{1.33}  & \multicolumn{1}{c|}{0.74}  & \multicolumn{1}{c|}{0.49}         & 0.36          \\ \cline{2-14} 
                      & 16                  & \multicolumn{1}{l|}{2,436}   & \multicolumn{1}{c|}{4,531}   & \multicolumn{1}{c|}{2,114}        & 1,247         & \multicolumn{1}{c|}{ 3.14} & \multicolumn{1}{c|}{5.6}   & \multicolumn{1}{c|}{7.9}          & 3.58          & \multicolumn{1}{c|}{2.42}  & \multicolumn{1}{c|}{0.75}  & \multicolumn{1}{c|}{0.5}          & 0.37          \\ \cline{2-14} 
\multirow{-3}{*}{8}   & 32                  & \multicolumn{1}{l|}{4,458}   & \multicolumn{1}{c|}{9,456}   & \multicolumn{1}{c|}{4,284}        & 1,627         & \multicolumn{1}{c|}{ 3.00} & \multicolumn{1}{c|}{10.0}  & \multicolumn{1}{c|}{10.5}         & 3.77          & \multicolumn{1}{c|}{4.28}  & \multicolumn{1}{c|}{0.77}  & \multicolumn{1}{c|}{0.52}         & 0.38          \\ \hline
                      & 8                   & \multicolumn{1}{l|}{3,155}   & \multicolumn{1}{c|}{4,511}   & \multicolumn{1}{c|}{2,288}        & 1,477         & \multicolumn{1}{c|}{ 3.35} & \multicolumn{1}{c|}{5.3}   & \multicolumn{1}{c|}{6.2}          & 4.15          & \multicolumn{1}{c|}{1.63}  & \multicolumn{1}{c|}{0.87}  & \multicolumn{1}{c|}{0.49}         & 0.36          \\ \cline{2-14} 
                      & 16                  & \multicolumn{1}{l|}{5,169}   & \multicolumn{1}{c|}{8,901}   & \multicolumn{1}{c|}{4,042}        & 2,133         & \multicolumn{1}{c|}{ 3.22} & \multicolumn{1}{c|}{8.9}   & \multicolumn{1}{c|}{9.6}          & 4.36          & \multicolumn{1}{c|}{2.98}  & \multicolumn{1}{c|}{0.89}  & \multicolumn{1}{c|}{0.5}          & 0.37          \\ \cline{2-14} 
\multirow{-3}{*}{16}  & 32                  & \multicolumn{1}{l|}{9,630}   & \multicolumn{1}{c|}{17,274}  & \multicolumn{1}{c|}{8,046}        & 2,894         & \multicolumn{1}{c|}{ 3.18} & \multicolumn{1}{c|}{15.9}  & \multicolumn{1}{c|}{13.0}         & 4.74          & \multicolumn{1}{c|}{5.22}  & \multicolumn{1}{c|}{0.93}  & \multicolumn{1}{c|}{0.52}         & 0.38          \\ \hline
                      & 8                   & \multicolumn{1}{l|}{5,546}   & \multicolumn{1}{c|}{9,235}   & \multicolumn{1}{c|}{4,079}        & 2,599         & \multicolumn{1}{c|}{ 3.51} & \multicolumn{1}{c|}{8.4}   & \multicolumn{1}{c|}{9.6}          & 6.41          & \multicolumn{1}{c|}{2.02}  & \multicolumn{1}{c|}{1.07}  & \multicolumn{1}{c|}{0.49}         & 0.36          \\ \cline{2-14} 
                      & 16                  & \multicolumn{1}{l|}{9,262}   & \multicolumn{1}{c|}{17,634}  & \multicolumn{1}{c|}{7,307}        & 3,913         & \multicolumn{1}{c|}{ 3.31} & \multicolumn{1}{c|}{13.8}  & \multicolumn{1}{c|}{15.0}         & 6.82          & \multicolumn{1}{c|}{3.92}  & \multicolumn{1}{c|}{1.13}  & \multicolumn{1}{c|}{0.5}          & 0.37          \\ \cline{2-14} 
\multirow{-3}{*}{32}  & 32                  & \multicolumn{1}{l|}{19,282}  & \multicolumn{1}{c|}{27,811}  & \multicolumn{1}{c|}{13,118}       & 5,433         & \multicolumn{1}{c|}{ 3.29} & \multicolumn{1}{c|}{25.4}  & \multicolumn{1}{c|}{20.8}         & 7.57          & \multicolumn{1}{c|}{6.07}  & \multicolumn{1}{c|}{1.12}  & \multicolumn{1}{c|}{0.52}         & 0.38          \\ \hline
                      & 8                   & \multicolumn{1}{l|}{10,093}  & \multicolumn{1}{c|}{19,028}  & \multicolumn{1}{c|}{7,706}        & 4,935         & \multicolumn{1}{c|}{ 3.80} & \multicolumn{1}{c|}{13.4}  & \multicolumn{1}{c|}{16.7}         & 10.92         & \multicolumn{1}{c|}{2.42}  & \multicolumn{1}{c|}{1.33}  & \multicolumn{1}{c|}{0.5}          & 0.37          \\ \cline{2-14} 
                      & 16                  & \multicolumn{1}{l|}{19,003}  & \multicolumn{1}{c|}{29,259}  & \multicolumn{1}{c|}{12,981}       & 7,563         & \multicolumn{1}{c|}{ 3.58} & \multicolumn{1}{c|}{22.5}  & \multicolumn{1}{c|}{26.1}         & 11.74         & \multicolumn{1}{c|}{4.02}  & \multicolumn{1}{c|}{1.35}  & \multicolumn{1}{c|}{0.51}         & 0.38          \\ \cline{2-14} 
\multirow{-3}{*}{64}  & 32                  & \multicolumn{1}{l|}{37,031}  & \multicolumn{1}{c|}{56,589}  & \multicolumn{1}{c|}{25,342}       & 10,604        & \multicolumn{1}{c|}{ 3.50} & \multicolumn{1}{c|}{41.2}  & \multicolumn{1}{c|}{37.0}         & 13.25         & \multicolumn{1}{c|}{6.92}  & \multicolumn{1}{c|}{1.37}  & \multicolumn{1}{c|}{0.53}         & 0.39          \\ \hline
                      & 8                   & \multicolumn{1}{l|}{19,417}  & \multicolumn{1}{c|}{33,916}  & \multicolumn{1}{c|}{15,945}       & 9,689         & \multicolumn{1}{c|}{ 3.34} & \multicolumn{1}{c|}{21.4}  & \multicolumn{1}{c|}{23.3}         & 15.11         & \multicolumn{1}{c|}{2.73}  & \multicolumn{1}{c|}{1.62}  & \multicolumn{1}{c|}{0.5}          & 0.37          \\ \cline{2-14} 
                      & 16                  & \multicolumn{1}{l|}{37,128}  & \multicolumn{1}{c|}{60,686}  & \multicolumn{1}{c|}{27,585}       & 14,945        & \multicolumn{1}{c|}{ 3.98} & \multicolumn{1}{c|}{37.1}  & \multicolumn{1}{c|}{40.2}         & 16.75         & \multicolumn{1}{c|}{4.73}  & \multicolumn{1}{c|}{1.63}  & \multicolumn{1}{c|}{0.51}         & 0.38          \\ \cline{2-14} 
\multirow{-3}{*}{128} & 32                  & \multicolumn{1}{l|}{84,657}  & \multicolumn{1}{c|}{115,835} & \multicolumn{1}{c|}{53,790}       & 21,028        & \multicolumn{1}{c|}{ 3.07} & \multicolumn{1}{c|}{69.1}  & \multicolumn{1}{c|}{57.1}         & 19.77         & \multicolumn{1}{c|}{8.04}  & \multicolumn{1}{c|}{1.63}  & \multicolumn{1}{c|}{0.53}         & 0.39          \\ \hline
                      & 8                   & \multicolumn{1}{l|}{39,155}  & \multicolumn{1}{c|}{74,719}  & \multicolumn{1}{c|}{31,693}       & 19,086        & \multicolumn{1}{c|}{ 4.48} & \multicolumn{1}{c|}{36.5}  & \multicolumn{1}{c|}{47.5}         & 29.87         & \multicolumn{1}{c|}{2.93}  & \multicolumn{1}{c|}{1.91}  & \multicolumn{1}{c|}{0.5}          & 0.37          \\ \cline{2-14} 
                      & 16                  & \multicolumn{1}{l|}{74,170}  & \multicolumn{1}{c|}{126,804} & \multicolumn{1}{c|}{55,682}       & 29,599        & \multicolumn{1}{c|}{ 3.87} & \multicolumn{1}{c|}{62.1}  & \multicolumn{1}{c|}{73.6}         & 33.15         & \multicolumn{1}{c|}{5.62}  & \multicolumn{1}{c|}{1.94}  & \multicolumn{1}{c|}{0.51}         & 0.38          \\ \cline{2-14} 
\multirow{-3}{*}{256} & 32                  & \multicolumn{1}{l|}{103,622} & \multicolumn{1}{c|}{234,957} & \multicolumn{1}{c|}{101,725}      & 41,763        & \multicolumn{1}{c|}{ 3.20} & \multicolumn{1}{c|}{113.0} & \multicolumn{1}{c|}{105.4}        & 39.20         & \multicolumn{1}{c|}{10.02} & \multicolumn{1}{c|}{1.97}  & \multicolumn{1}{c|}{0.53}         & 0.39          \\ \hline
\end{tabular}

\vspace{-0.5em}
\end{table*}

\ignore{
\begin{figure*}[]
\vspace{-0.7em}
     \centering
     \begin{subfigure}[]{ 0.30\textwidth}
         \centering
         \includegraphics[width=6cm]{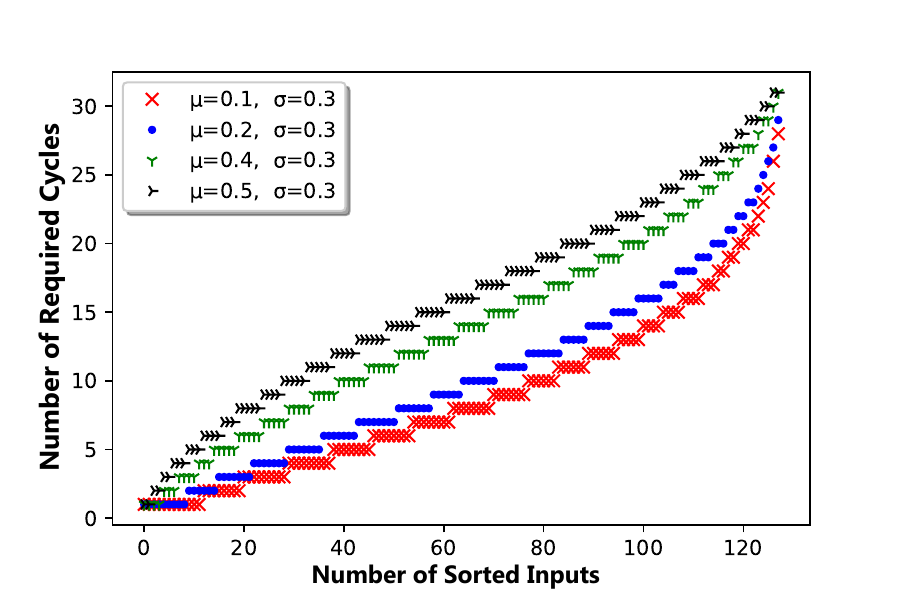}
         \caption{$M=5$}
         \label{fig21}
     \end{subfigure}
     \begin{subfigure}[]{0.30\textwidth}
         \centering
         \includegraphics[width=6cm]{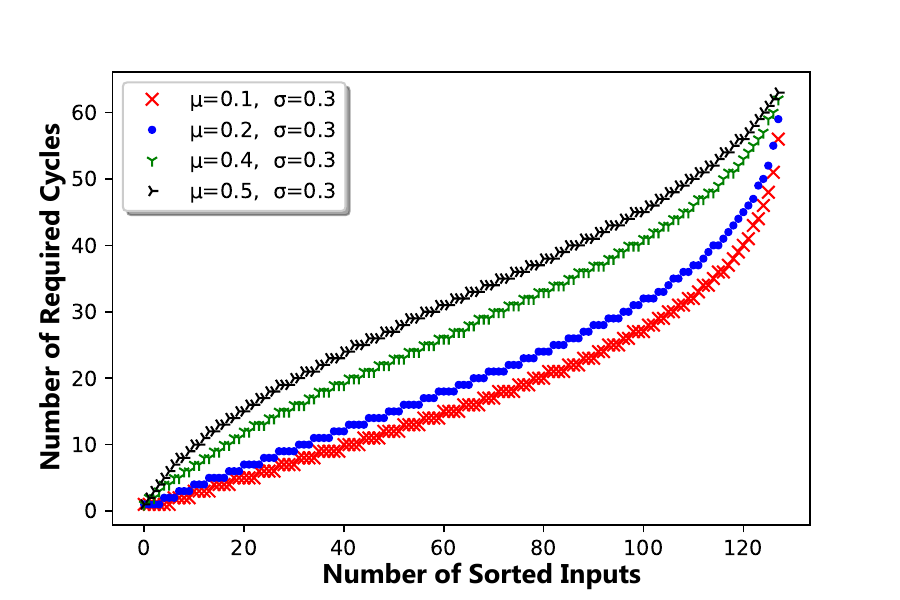}
         \caption{$M=6$}
         \label{fig22}
     \end{subfigure}
     \begin{subfigure}[]{ 0.30\textwidth}
         \centering
         \includegraphics[width=6cm]{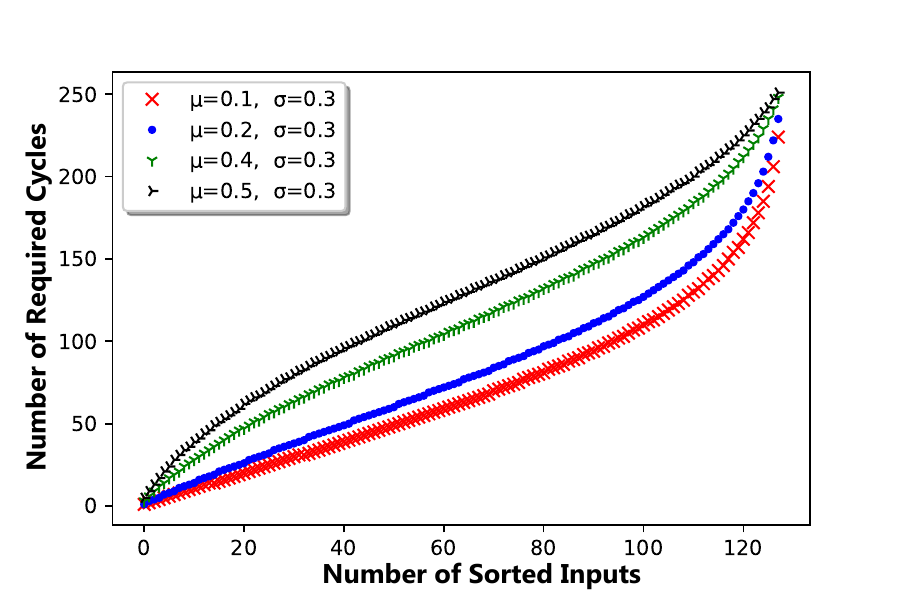}
         \caption{$M=8$}
         \label{fig23}
     \end{subfigure}

     \hfill

     \begin{subfigure}[b]{height = 0.3\textwidth}
         \centering
         \includegraphics[width=5cm]{M=8}
         \caption{$M=8$}
         \label{fig24}
     \end{subfigure}

\vspace{-0.5em}
\caption{Number of 
processing cycles to iteratively  find the minimum values for data with $M=5,6$ and $8$ bit-widths. 
  }
        \label{fig2}
\end{figure*}
}

\begin{figure*}[]
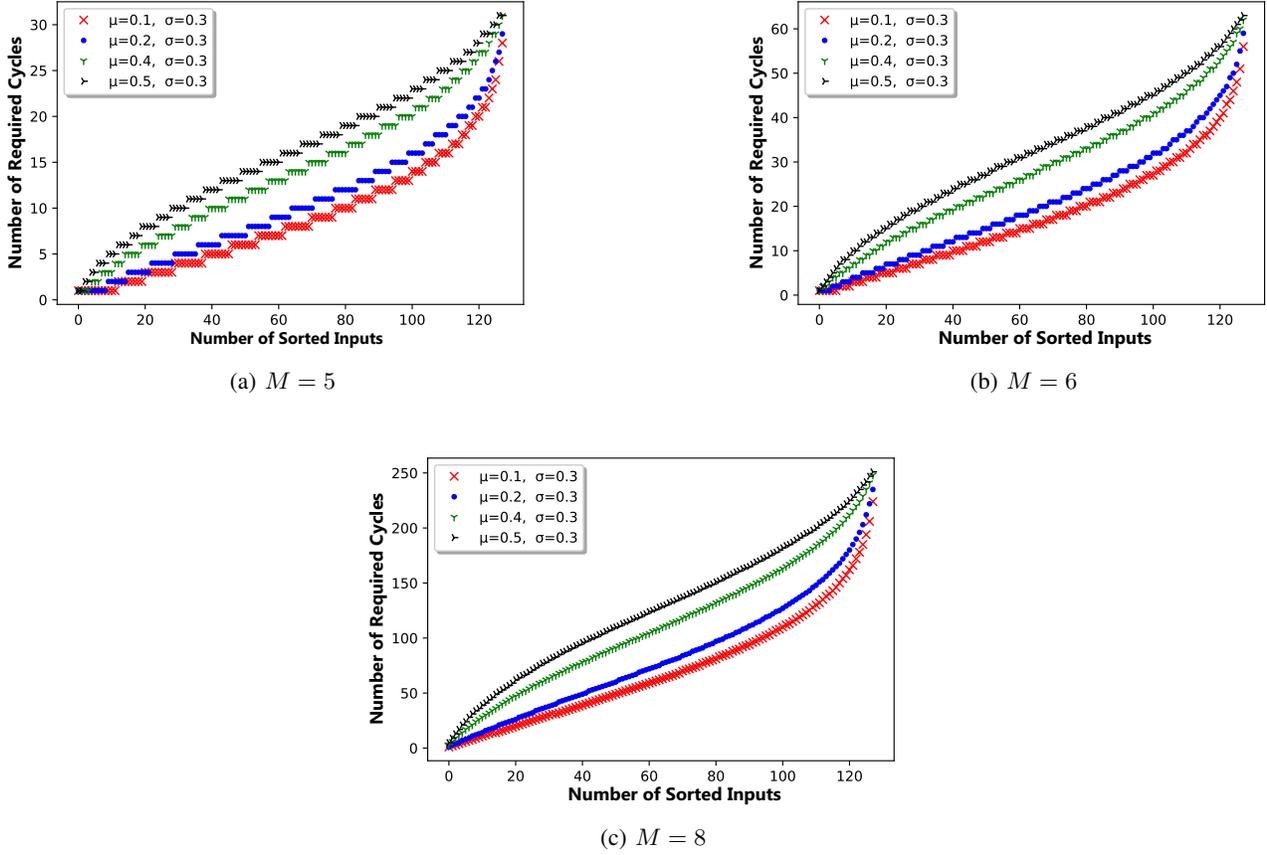

\centering
\begin{subfigure}[]{0.45\textwidth}
    \centering
    \includegraphics[width=8cm]{M=5}
    \caption{$M=5$}
    \label{fig21}
\end{subfigure}
\hfill
\begin{subfigure}[]{0.45\textwidth}
    \centering
    \includegraphics[width=8cm]{M=6}
    \caption{$M=6$}
    \label{fig22}
\end{subfigure}
\vspace{0.5em} 

\begin{subfigure}[]{0.45\textwidth}
    \centering
    \includegraphics[width=8cm]{M=8}
    \caption{$M=8$}
    \label{fig23}
\end{subfigure}
\vspace{-0.5em}
\caption{Number of processing cycles to iteratively find the minimum values for data with $M=5,6$ and $8$ bit-widths.}
\label{fig2}
\end{figure*}

\textit{SED} finds the index of the bit-stream with 
minimum value. This is done by finding the bit-stream that produces the first 0.\ignore{largest element detector which includes a left-aligned unary number generator and the largest element detector.} Consider a set of 
three inputs, $p_1=0.5$, $p_2=0.75$, and $p_3=0.5$. 
A right-aligned unary representation for these numbers with $M$=3 is 
$p_1=00001111$, $p_2=00111111$, and $p_3=00001111$. 
In the fifth cycle, a zero is generated for $p_1$ 
(the blue '0' in \ding{203}) and the same for $p_3$.
This enables the flip-flops corresponding to the first (the green '1' in \ding{204}) and third inputs.
When these 
flip-flops are activated, the detection signal (\textit{ds}), which is the output of an \textit{addition} unit, will have a value of two. \textit{ds} $=2$ means that the next minimum
value is not a single number but two numbers with the same value.
We utilize a \textit{priority encoder} to obtain the memory address of one of the minimum values in the fifth cycle. Next, \textit{ds} is passed to a \textit{controller}. Fig.~\ref{Controller} shows the controller's FSM. 
When $ds=2$, the state changes from "Find the index" to "Put the results." The state does not change until the two numbers are at the output. Then, the \textit{controller} enables a \textit{down counter} that gives 
the address of the 
numbers found in the "Put the result" state. 
A \textit{MUX} unit is also used to get the address of the minimum values from the sorting module  and put the weighted-binary value of the maximum on the output registers. Since \Design subtracts the input value from the last generated bit in the unary stream, the input memory of $p_1$ and $p_3$ are set to zero in the fifth cycle. To 
retrieve the correct value of the input data,  we need to add the current value of $p_1$ (which is 0)  with  (\textit{Elapsed\_Cycle}$-1$)
from the controller.  Consequently, the value of $0.5$ is retrieved and stored in the output memory. 

\section{Design Evaluation}
\label{Design_Eval}

To rigorously evaluate the proposed comparison-free unary sorting architecture, we implemented RTL-level models in VHDL and synthesized them using Synopsys Design Compiler (v2018.06) targeting a 45\,nm FreePDK standard-cell library. The evaluation spanned multiple configurations with varying input sizes ($N = 8$ to $256$) and data bit-widths ($M = 8$, $16$, and $32$). The performance of the proposed design was benchmarked against three prominent state-of-the-art designs: the unary comparison-free sorter with traditional comparator-based UNG~\cite{AmirDAC2022}, the general-purpose comparison-free sorting engine~\cite{Ghosh2019}, and the UC-based Batcher sorting network~\cite{Najafi2018Sorting}. Table~\ref{tbl:2} presents the results in terms of area, power, and critical path delay.

In the representative configuration of $N = 32$ and $M = 16$, the proposed design achieves an area of $3,913\,\mu m^2$, compared to $7,307\,\mu m^2$ in~\cite{AmirDAC2022}, $17,634\,\mu m^2$ in~\cite{Najafi2018Sorting}, and $9,262\,\mu m^2$ in~\cite{Ghosh2019}. This indicates area reductions of approximately 46.4\%, 77.8\%, and 57.7\%, respectively. Similarly, power consumption in this setting is $6.82$\,mW, while the other designs consume $15.0$\,mW~\cite{AmirDAC2022}, $13.8$\,mW~\cite{Najafi2018Sorting}, and $3.31$\,mW~\cite{Ghosh2019}. 
The proposed sorter achieves a critical path delay of $0.37$\,ns, significantly shorter than $0.50$\,ns, $1.13$\,ns, and $3.92$\,ns in the respective baseline designs.

These trends hold consistently as the input size increases. For example, at $N = 128$ and $M = 32$, the proposed sorter occupies only $21,028\,\mu m^2$, compared to $53,790\,\mu m^2$~\cite{AmirDAC2022}, $115,835\,\mu m^2$~\cite{Najafi2018Sorting}, and $84,657\,\mu m^2$~\cite{Ghosh2019}, offering area savings of more than 60\% in all cases. Power consumption is reduced to $19.77$\,mW, outperforming $37.0$\,mW~\cite{AmirDAC2022} and $57.1$\,mW~\cite{Najafi2018Sorting}, while the critical path delay remains only $0.39$\,ns, far lower than the $0.53$\,ns, $1.63$\,ns, and $8.04$\,ns observed in the other designs.

Beyond static synthesis metrics, we also evaluated runtime efficiency. The proposed architecture sorts data by iteratively detecting and extracting the minimum value at each cycle, allowing partially sorted outputs to emerge early. This results in lower overall cycle counts for complete sorting. Fig.~\ref{fig2} shows the number of clock cycles required to identify the $n^\text{th}$ minimum value for input data sampled from Gaussian distributions $X \sim \mathcal{N}(\mu, \sigma^2)$. The results confirm that the proposed design consistently requires fewer cycles to reach convergence, particularly under distributions with low variance, which further demonstrates its algorithmic efficiency and low sensitivity to input randomness.

In summary, the proposed sorting architecture achieves up to 82\% reduction in area, up to 70\% reduction in power consumption, and up to 92\% reduction in critical path delay compared to state-of-the-art designs. These advantages make it a compelling solution for hardware sorting in energy- and area-constrained environments, such as embedded systems and real-time computing platforms.

\section{CONCLUSION}
\label{Conclusion}

This work presented a power- and area-efficient comparison-free unary sorting architecture based on a novel finite-state-machine-based unary number generator (CFUNG). By replacing the traditional comparator-based unary generation unit with a compact FSM design, the proposed sorter significantly reduces the cost of unary stream generation—previously one of the major bottlenecks in unary computing-based sorting modules.
The proposed architecture performs ascending-order sorting by iteratively identifying minimum values using right-aligned unary bitstreams. Synthesis results, obtained across a wide range of configurations ($N = 8$ to $256$, $M = 8$ to $32$), demonstrate that the proposed design achieves up to 82\% reduction in area and 70\% reduction in power consumption compared to the state-of-the-art design in~\cite{AmirDAC2022}. Furthermore, the proposed architecture exhibits superior timing characteristics, with critical path delays reduced by over 90\% in large-scale configurations.

In addition to static performance metrics, the design also shows improved runtime efficiency due to its iterative sorting mechanism, which reduces the number of clock cycles required to reach fully sorted outputs. These improvements collectively establish the proposed sorter as a highly viable and scalable solution for energy-constrained and area-sensitive applications, particularly in embedded and real-time computing environments.

\bibliographystyle{IEEEtran}
\bibliography{sample-base}

\end{document}